\documentclass[aps,pra,twocolumn,showpacs,superscriptaddress,longbibliography]{revtex4-1} 
\usepackage{graphicx}
\usepackage{amsmath}
\usepackage{amssymb}
\usepackage{epstopdf}
\usepackage{amsfonts}
\usepackage{dcolumn}
\usepackage{multirow}

\begin{document}
\title{Broken scale-invariance in time-dependent trapping potentials}
\author{Seyed Ebrahim Gharashi}
\affiliation{Department of Physics and Astronomy,
Washington State University,
  Pullman, Washington 99164-2814, USA}
\author{D. Blume}
\affiliation{Department of Physics and Astronomy,
Washington State University,
  Pullman, Washington 99164-2814, USA}
\date{\today}

\begin{abstract}
The response of a cold atom gas with contact interactions
to a smoothly varying external harmonic confinement in
the non-adiabatic regime is studied.
The time variation of the angular frequency is varied
such that the system is, for vanishing or
infinitely strong contact interactions, scale invariant.
The time evolution of the system with broken scale invariance 
(i.e., the time evolution of the system with finite interaction strength),
is contrasted with that for a scale invariant system,
which exhibits Efimovian-like expansion dynamics
that is characterized by log-periodic oscillations with unique
period and amplitude.
It is found that the breaking of the scale invariance by the 
finiteness of the interactions leads to a time dependence of the
oscillation period and amplitude.
It is argued, based on analytical 
considerations 
for atomic gases of arbitrary size
and numerical results for two one-dimensional particles, that the oscillation period
approaches that of the scale-invariant system
at large times. The role of the time-dependent contact in the
expansion dynamics is analyzed.

\end{abstract}
\maketitle

\section{Introduction}
\label{sec_introduction}

The response of a system to a perturbation lies at the 
heart of important concepts such as linear response theory~\cite{Fetter03}, 
characterizing whether a system is chaotic or not~\cite{Stockmann06}, 
state engineering~\cite{Nikolopoulos13,Menchon-Enrich16}, 
and adiabatic transport~\cite{Polo16}.
This paper considers the quantum mechanical response of an 
initial state to a time-dependent variation of the system Hamiltonian. 
In the context of cold atom 
systems~\cite{Polkovnikov11,Yan13,Aoki14,Makotyn14,Langen15,Will15,Cetina16,Blakie04,Gericke07,Rosi13,Masuda14},
two limiting cases have received considerable attention, 
the regime where the system Hamiltonian is changed 
adiabatically and the regime where the system Hamiltonian is quenched, 
i.e., changed significantly over a time scale 
that is short compared to the intrinsic 
time scales of the system (essentially instantaneously).
The time dynamics of small cold atom systems has specifically attracted
a great deal of attention recently, either because the few-body dynamics
is interesting in its own right or as a model for understanding intricate
many-body
dynamics~\cite{Lode12,Sykes14,Zurn12,Zurn13,Gharashi15-tunn,Ebert15,Corson16}.
Here, we consider an ``in between'' 
case in which the system Hamiltonian is changed continuously on a time scale that is comparable to the intrinsic time scale of the system Hamiltonian at time $t_0$,
where
$t_0$ is the time at which the time variation of the Hamiltonian is 
turned on.
Starting with an eigenstate at $t_0$, 
we focus on the regime where the long time dynamics exhibits 
oscillations.

Reference~\cite{Deng16} discussed an intriguing analogy between the dynamics of an $N$-atom system with vanishing or infinitely large two-body zero-range interactions under time-dependent external harmonic confinement and the static three-body Efimov solution~\cite{Efimov70,Braaten06}. 
The former exhibits, for a properly chosen time variation of the trap, log-periodic expansion dynamics~\cite{Deng16} 
while the latter is, for infinitely large $s$-wave scattering length, characterized by log-periodic energy spacings~\cite{Efimov70}.
It was shown~\cite{Deng16} 
that the log-periodic time evolution can be traced back to the scale invariance of the $N$-body Hamiltonian.
Specifically, by changing the angular trap frequency with time, the harmonic oscillator length can be effectively removed from the problem, leaving a scale-invariant space-time Hamiltonian that is governed by intriguing long-term dynamics.
The present paper investigates how the dynamics changes when
the two-body interactions define a length scale.
How does the time evolution change when the interaction strength increases from zero to infinity?
Do the long time oscillations survive for finite interaction strengths?
Can one still define an oscillation period?

The remainder of this paper is organized as follows.
Section~\ref{sec_theory} introduces 
the system Hamiltonian and theoretical background,
with Secs.~\ref{sec_scale-inv} and~\ref{sec_broken_scale-inv} 
discussing the cases where the two-body interactions, respectively, 
do not and do
define a meaningful length scale.
Section~\ref{sec_results} presents our 
numerical results for a two-particle system and discusses the findings.
Last, Sec.~\ref{sec_conclusion} summarizes.


\section{System Hamiltonian and theoretical background}
\label{sec_theory}

\subsection{General considerations}
\label{sec_theory_intro}
We
consider $N$ atoms with mass $m$ and position vectors ${\mathbf{r}}_j$
($j=1,\dots,N$) under external isotropic harmonic confinement with time-dependent angular 
trapping frequency $\omega(t)$ and 
two-body zero-range interactions 
$g \delta^{(d)}({\mathbf{r}}_j - {\mathbf{r}}_k)$ 
between each pair of particles $j$ and $k$.
The coupling constant $g$, which has units of 
$ energy \cdot length^d$, 
can be tuned in cold atom experiments via 
Feshbach resonance techniques~\cite{Chin10}.
In the non-interacting limit ($g=0$) and the 
infinitely strongly-interacting limit ($g=\infty$), 
the two-body interactions do not define a meaningful length 
scale~\cite{Giorgini08,Zwerger11} and 
the hyperradial and hyperangular degrees of freedom 
decouple~\cite{Jonsell02,Werner06PRA,footnote7-1}.
In this case, the one-dimensional 
Schr\"odinger-like equation for the hyperradius $R$,
\begin{equation}
R = \sqrt{ \sum_{j=1}^N \frac{{\mathbf{r}}_j^2}{N}},
\label{Eq_hyperradius}
\end{equation}
takes a particularly simple form,
\begin{align}
\left[
-\frac{\hbar^2}{2 N m} \frac{\partial^2}{\partial R^2}
+\frac{1}{2} N m \omega^2(t)  R^2
+\frac{\hbar^2 \mathcal{C}}{2 N m R^2}
\right]
\psi(R,t)
=
\nonumber \\
E \psi(R,t),
\label{Eq_Sch-hyperradius}
\end{align}
where ${\mathcal{C}}$ is a constant
that is determined by the 
eigenvalue of the, in general, $(d N -1)$-dimensional
differential equation
for the hyperangular degrees of freedom
(at this point, we do not separate off the $d$ center of mass degrees
of freedom).
While determining ${\mathcal{C}}$ 
for $g = \infty$ is, in general (for $N>3$), 
a highly non-trivial 
task~\cite{Werner06PRA,vonStecher08,vonStecher09CG}, 
a crucial point is that the $N$-body dynamics is, 
for $g=0$ and $g=\infty$, 
fully governed by the one-dimensional Schr\"odinger-like equation 
for the hyperradius $R$.
The solutions to 
Eq.~(\ref{Eq_Sch-hyperradius}) and hence of the $N$-body system with zero-range interactions of vanishing or infinite strength have been discussed extensively in the literature since the 60s~\cite{Lewis67,Lewis68,Lewis69,Popov69,Popov70,Camiz71,Castin04,Moroz12,Ebert15}.

In what follows we assume $\omega(t)=\Omega$ for $t \le t_0$, where
$\Omega$ is a (real) constant. 
The solution to Eq.~(\ref{Eq_Sch-hyperradius}) 
can then be reduced to solving the differential equation
\begin{equation}
\ddot{\Lambda}(t) - 
\frac{\Omega^2}{\Lambda^3(t)} + \omega^2(t) \Lambda(t)=0,
\label{Eq_Lambda-diff}
\end{equation}
where $\Lambda(t)$ is a scaling function.
Assuming the system is initially, at $t=t_0$,
in the eigenstate $\psi_{\rm{eigen}}(R,t_0)$, the 
wave packet $\psi(R,t)$ for $t>t_0$ is given by~\cite{Popov69,Popov70}
\begin{equation}
\psi\left(R,t\right) =
{\mathcal{N}}(t) 
\exp\left( \frac{i m \dot{\Lambda}(t) N R^2}{2 \hbar \Lambda(t)} \right)
\psi_{\rm{eigen}}\left(\frac{R}{\Lambda(t)},t_0\right),
\label{Eq_WP}
\end{equation}
where $\Lambda (t)$ is a solution to Eq.~(\ref{Eq_Lambda-diff}) 
with 
$\Lambda(t_0) = 1$ and $\dot{\Lambda}(t_0)=\ddot{\Lambda}(t_0)=0$.
The time-dependent normalization factor
${\mathcal{N}}(t)$ reads~\cite{Popov69,Popov70,Castin04}
\begin{equation}
{\mathcal{N}} (t) = [\Lambda(t)]^{-d N/2}
\exp\left(- i \frac{E}{\hbar} \int_{t_0}^t
\frac{d t'}{\Lambda^2(t')}\right).
\end{equation}
It should be noted that the constant ${\mathcal{C}}$ does
not enter into the differential equation for the scaling 
function $\Lambda(t)$; it enters into the 
wave packet $\psi(R,t)$
for $t>t_0$ solely through the functional form 
of the eigenstate 
$\psi_{\rm{eigen}}(R,t_0)$ at $t=t_0$.

Importantly, the hyperangular degrees of 
freedom are not affected by the time variation of $\omega(t)$.
This implies that the hyperangular portion of the 
wave function is stationary and that the 
full wave packet $\Psi({\mathbf{R}},t)$, 
i.e., the wave packet that depends on the hyperradius $R$ and 
hyperangles $\hat{\mathbf{R}}$, where
${\mathbf{R}}$ collectively denotes all the 
position vectors, i.e., 
${\mathbf{R}}=({\mathbf{r}}_1,\cdots,{\mathbf{r}}_N)$,
can be readily constructed from Eq.~(\ref{Eq_WP}), 
provided the hyperangular portion of the wave function is known at $t=t_0$.

Historically~\cite{Lewis67,Lewis68,Lewis69}, 
the wave packet dynamics of Eq.~(\ref{Eq_Sch-hyperradius})
has been introduced and analyzed in connection with the
corresponding classical harmonic oscillator equation
for the generalized coordinate $\eta(t)$,
\begin{eqnarray}
\label{eq_classicalho}
\ddot{\eta}(t) + \omega^2(t) \eta(t)=0,
\end{eqnarray}
where, as in the quantum case, $\omega(t)=\Omega$ for $t \le t_0$.
Seeking complex solutions of the form 
$\eta(t) = \Lambda(t) \exp[\imath \gamma(t)]$,
one obtains coupled differential equations for $\Lambda(t)$ 
and $\gamma(t)$.
The differential equation for $\Lambda(t)$ is, for
the initial conditions $\Lambda(t_0)=1$ and $\dot{\gamma}(t_0)=-\Omega$,
independent of $\gamma(t)$
and identical to that given in Eq.~(\ref{Eq_Lambda-diff}).
This correspondence between the absolute value $|\eta(t)|$ of the
classical generalized coordinate 
and the 
quantum mechanical scaling function $\Lambda(t)$
is used in Sec.~\ref{sec_scale-inv} 
to elucidate some characteristics
of the quantum mechanical $N$-body problem in which the two-body 
interactions do not define a meaningful length scale.

If the coupling constant $g$ is finite, 
the hyperradial and hyperangular degrees of freedom are,
in general, coupled. 
This implies that 
the wave packet dynamics depends, in general, on all $d N$ degrees
of freedom; it cannot be reduced to a one-dimensional
problem.
We note, however, that the center of mass degrees of freedom
can be separated off
(see Sec.~\ref{sec_results} for more details).
The coupling between the hyperangular
and hyperradial degrees of freedom
also implies that the quantum-classical correspondence
breaks down.

In the remainder of this paper 
we consider a time variation of $\omega(t)$
for which the time-dependent harmonic oscillator length
$a_{\rm{ho}}(t)$, where 
$a_{\rm{ho}} (t) = \sqrt{\hbar / [m \omega(t)]}$, decreases as 
$1/\sqrt{t}$ for $t>t_0$, i.e., we consider
\begin{equation}
\omega (t) = \Bigg\{
\begin{array} {l} 
\Omega \mbox{ for } t \le t_0 \\
\Omega  \frac{t_0}{t} \mbox{ for } t > t_0 
\end{array}.
\label{Eq_omega}
\end{equation}
If the angular frequency $\omega(t)$ changes little 
when $t$ increases from $t_0$ to $t_0+T$, where $T$
is the characteristic time scale of the non-interacting system
for $t_0 \le t$ ($T = 2 \pi /\Omega$),
the resulting dynamics is adiabatic. This situation is
realized when $\Omega t_0$ is much larger than 1.
In what follows, we 
primarily look at time variations characterized
by ``medium'' speeds, i.e.,
situations for which $\Omega t_0$ is not large compared to $1$
(our numerical calculations in Sec.~\ref{sec_results}, e.g., use
$\Omega t_0=\sqrt{10}$).

Section~\ref{sec_scale-inv} discusses the $N$-particle Hamiltonian
with scale invariant 
interactions, which was the subject of Ref.~\cite{Deng16}, 
and Sec.~\ref{sec_broken_scale-inv} 
considers interactions that define a scale
(i.e., interactions with finite $g$).

\subsection{Scale invariant interactions}
\label{sec_scale-inv}


This section focuses on the $g=0$ and $1/g=0$ cases.
Using Eq.~(\ref{Eq_WP}), the 
expectation values $\langle R^n(t) \rangle$ 
can be calculated analytically for any positive integer $n$.
We find that these expectation values are fully determined by
the scaling function $\Lambda(t)$ and
the initial expectation value $\langle R^n(t_0) \rangle$,
\begin{equation}
\label{eq_exp_hyper}
 \frac{\langle R^n(t)\rangle}{\langle R^n(t_0)\rangle}  = 
  \Lambda^n(t).
\end{equation}
The $n=2$ expression was used in Ref.~\cite{Deng16} to analyze the
experimental expansion images.
Equation~(\ref{eq_exp_hyper}) implies that the 
quantum mechanical
hyperradial motion can be interpreted
using the correspondence with the
time-dependent classical harmonic oscillator,
i.e.,
the expectation value $\langle R(t) \rangle$ and
its
time variation $d(\langle R(t) \rangle)/dt$ can be visualized in 
``phase space'' by plotting $\dot{\Lambda}(t)$ as a function of
$\Lambda(t)$.
More generally,
the
expectation value $\langle R^n(t) \rangle$ and
its
time variation $d(\langle R^n(t) \rangle)/dt$ can be visualized in 
phase space by plotting $d \Lambda^n(t)/dt$ as a function of
$\Lambda^n(t)$.

The solution $\Lambda(t)$ to 
Eq.~(\ref{Eq_Lambda-diff}) for the time-dependent
potential given in Eq.~(\ref{Eq_omega})
has distinct functional forms for $\lambda>4$ and
$\lambda<4$, where
$\lambda$ is defined as $(\Omega t_0)^{-2}$.
Using as before $\Lambda(t_0)=1$ and
$\dot{\Lambda}(t_0)=\ddot{\Lambda}(t_0)=0$,
the solution for $\lambda>4$ reads~\cite{Lewis68}
\begin{eqnarray}
\Lambda(t)= \left\{
\frac{t \left(\eta^2 -1\right)}{t_0 \eta^2}
\left[
1-\frac{1}{2}
\left(
\frac{(t/t_0)^{\eta}}{\eta+1}
-\frac{(t/t_0)^{-\eta}}{\eta-1}
\right)
\right]
\right\}^{1/2},
\end{eqnarray}
where $\eta = \sqrt{1 - 4/\lambda}$, and that 
for $\lambda<4$ reads~\cite{Lewis68}
\begin{equation}
\Lambda (t) = \left\{
\frac{t}{t_0 \sin^2 (\varphi)}
\left[1-\cos \varphi \cos \left(s_0 \ln\left(\frac{t}{t_0}\right) + 
\varphi \right)\right]
\right\}^{1/2},
\label{Eq_sol_Zhai}
\end{equation}
where $s_0 = \sqrt{4/\lambda - 1}$ and $\varphi = - \arctan(s_0)$.
Reference~\cite{Deng16} made the nice observation that the $\lambda<4$ 
solution exhibits log-periodic behavior reminiscent of 
and formally equivalent to the
solutions to the static three-body Efimov problem.
In fact, the symbol $s_0$ is introduced to make the
connection to the static three-body Efimov scenario more 
explicit~\cite{Deng16}.

\begin{figure}[htbp]
\centering
\includegraphics[angle=0,width=0.4\textwidth,clip=true]{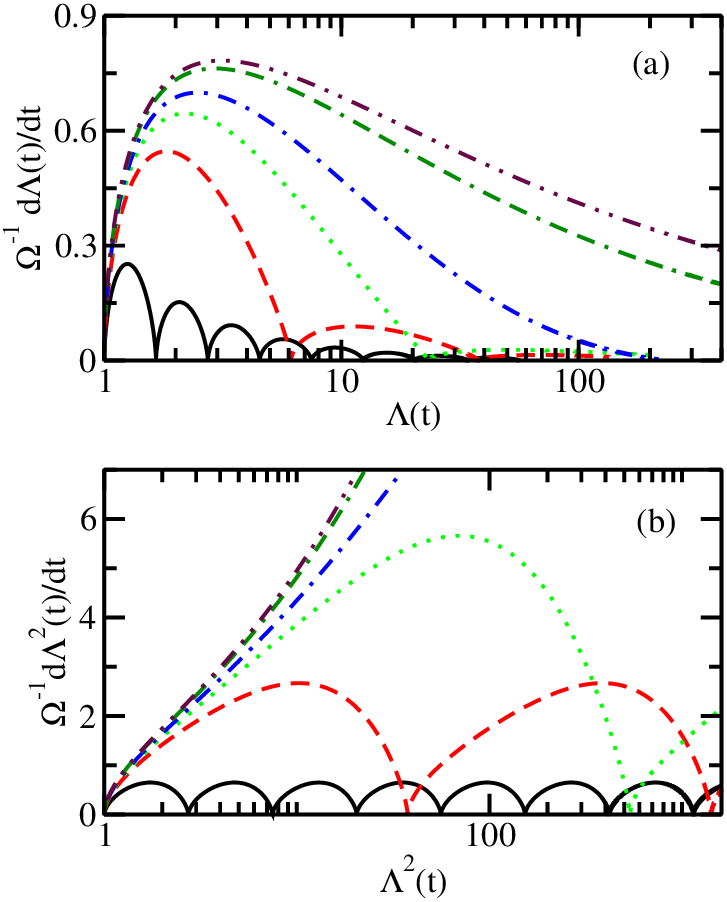}
\caption{(Color online)
Phase space trajectories that characterize the 
dynamics of the $N$-body system with scale invariant 
two-body interactions.
The lines show the
quantity $d \Lambda^n(t)/dt$ as a function of $\Lambda^n(t)$
for (a) $n=1$ and (b) $n=2$ for various $\lambda$.
The solid, dashed, dotted, dash-dotted, 
dash-dash-dotted, and dot-dot-dashed lines 
correspond to $\lambda= 0.1, 1,2,3,5,$ and $6$, respectively.
}
\label{fig_phasespace}
\end{figure}

Figure~\ref{fig_phasespace}(a) shows $\dot{\Lambda}(t)$ 
as a function of
$\Lambda(t)$ for various $\lambda$.
This phase space representation illustrates that the cloud size
increases continuously 
with increasing time. The 
expansion speed is greater or equal to zero for all $t$.
For $\lambda<4$
[solid, dashed, dotted, and dash-dotted lines
in Fig.~\ref{fig_phasespace}(a)],
the phase space trajectories 
for fixed $\lambda$ resemble those of a ``bouncing ball''. 
The value of $\lambda$,
assuming $\lambda<4$, does not affect the overall shape of the
trajectory; it merely changes the spacing between and amplitude
of the ``bounces'' (both the spacing and amplitude
decrease with decreasing $\lambda$).
For $\lambda>4$
[dash-dash-dotted and 
dot-dot-dashed lines in Fig.~\ref{fig_phasespace}(a)], 
the phase space trajectories show a finite positive
expansion speed for all $t > t_0$.
In these cases, the expansion speed first increases, then 
reaches a maximum, and eventually decreases monotonically
for all later times.
In the limit $\lambda \to \infty$,
i.e., in the limit of the
sudden removal of the trap, the expansion speed increases to 
a maximum that is set by the initial energy of the systems and 
then remains
constant after that.
Figure~\ref{fig_phasespace}(b) shows $d \Lambda^2(t)/dt$ 
as a function of
$\Lambda^2(t)$
for various $\lambda$. 
These phase space trajectories characterize,
according to Eq.~(\ref{eq_exp_hyper}),
the time dynamics of the expectation value $\langle R^2(t) \rangle$.
Qualitatively, Figs.~\ref{fig_phasespace}(a)
and \ref{fig_phasespace}(b) display the same characteristics.

To prepare for the discussion presented in 
the next section,
it is instructive to explicitly write down, as was done
in Ref.~\cite{Deng16}, the
differential equation for the 
expectation value 
$\langle R^2(t) \rangle$ of the square of the hyperradius,
\begin{equation}
\frac{d^3}{d t^3}\langle R^2(t)\rangle
+\frac{4}{\lambda t^2} \frac{d}{d t} \langle R^2(t)\rangle
-\frac{4}{\lambda t^3} \langle R^2(t)\rangle = 0.
\label{Eq_diff_Zhai}
\end{equation}
This equation
can be derived by applying 
Heisenberg's equation of 
motion to the $N$-particle Schr\"odinger
equation and taking advantage of the fact that the
two-body interactions do not define a meaningful 
length scale~\cite{Deng16}.
The next section considers how the differential 
equation for $\langle R^2(t) \rangle$,
Eq.~(\ref{Eq_diff_Zhai}), is modified
if the two-body interactions define a length scale.

\subsection{Interactions with finite $g$}
\label{sec_broken_scale-inv}

As already aluded to, the hyperradial and hyperangular degrees 
do, in general,  not decouple if the interaction strength $g$ of
the two-body interactions
is finite, making it challenging to tackle the full
$N$-body dynamics. 
To gain insight into the dynamics
for finite $g$, we consider the simplest
possible scenario, namely two
interacting particles in one spatial dimension.
In this case, the wave packet dynamics can be
determined numerically utilizing the 
techniques introduced in Ref.~\cite{Gharashi15-tunn}.
Moreover, 
analytical results for limiting cases can be used to
interpret the numerical results.

For the two-particle system,
the square
 of the hyperradius can be 
rewritten in terms of the relative coordinate
$z$ and the center of coordinate $Z_{\rm{CM}}$,
 $R^2=z^2/4 + (Z_{\rm{CM}})^2$ 
with $z = z_1-z_2$ and $Z_{\rm{CM}} = (z_1+z_2)/2$
(from now on
we use $z_j$ instead of $\mathbf{r}_j$ to emphasize
the one-dimensional nature of the system under study).
For the time variation defined
in Eq.~(\ref{Eq_omega}), the relative and center of mass degrees of 
freedom decouple, i.e., the full wave packet
$\Psi(z_1,z_2,t)$
can be written as a product of 
the relative part $\psi(z,t)$ and the center
of mass part $\Psi_{\rm{CM}}(Z_{\rm{CM}},t)$,
$\Psi(z_1,z_2,t)=\psi(z,t)\Psi_{\rm{CM}}(Z_{\rm{CM}},t)$.
Correspondingly, 
$\langle R^2(t)\rangle$ can be written as 
$\langle z^2(t)\rangle/4+\langle (Z_{\rm{CM}}(t))^2\rangle$.
This implies that the time evolution of the
relative and
center of mass
parts can be treated separately.
This also means that the two-body system considered does not
allow us to study the coupling between the hyperradial and hyperangular
degrees of freedom.
Nevertheless, it does allow us to investigate how the
finite interaction strength $g$ enters into the dynamics
of $\langle z^2(t) \rangle$.

The differential equation
for the center of mass part 
$\langle (Z_{\rm{CM}}(t))^2 \rangle$ 
is given by Eq.~(\ref{Eq_diff_Zhai})
with $R$ replaced by $Z_{\rm{CM}}$. Hence, the time evolution
of $\langle (Z_{\rm{CM}}(t))^2\rangle$ is the same as that for a single
(non-interacting) particle of mass $2m$.
The 
differential equation for the relative part
$\langle z^2(t) \rangle$ reads
\begin{equation}
\frac{d^3}{d t^3}\langle z^2(t)\rangle
+\frac{4}{\lambda t^2} \frac{d}{d t} \langle z^2(t)\rangle
-\frac{4}{\lambda t^3} \langle z^2(t)\rangle
=
-\frac{4 \hbar^2}{m^3 g} \frac{d C (t)}{d t},
\label{Eq_diff_1D}
\end{equation}
where the Tan contact $C(t)$ is defined through~\cite{Barth11}
\begin{equation}
C (t) = 
4 \hbar^2 \left[
\left|\frac{d \psi(z,t)}{d z}\right|^2 \right]_{z \to 0}.
\end{equation}
Noticing that the left hand sides of
Eqs.~(\ref{Eq_diff_Zhai}) and (\ref{Eq_diff_1D})
have the same functional
forms,
we conclude that 
the finiteness of the interaction strength $g$
enters into the differential equation for $\langle z^2(t)\rangle$
via the time variation of the Tan contact.
In fact, the right hand side of Eq.~(\ref{Eq_diff_1D})
vanishes for $g=0$ (in this case, the contact vanishes)
and $|g|=\infty$ (in this case, $1/|g|$ vanishes)
and Eq.~(\ref{Eq_diff_1D}) reduces to Eq.~(\ref{Eq_diff_Zhai})
in these cases.
Equation~(\ref{Eq_diff_1D}) was first
introduced for the general $N$ case in Refs.~\cite{Qi16,Shi16}.

To obtain a sense of how the contact $C(t)$ changes with time,
we imagine that the system changes adiabatically, i.e., we
determine the contact $C^{\rm{adia}}$
for the two-particle system
with coupling constant $g$
in a static harmonic trap with angular 
frequency $\bar{\omega}$ and
corresponding harmonic oscillator length $\bar{a}$.
Using the expressions for the
eigenenergies and eigenstates
from Ref.~\cite{Busch98}, the adiabatic contact for the
two-body system can be calculated
readily for any $g$.
In preparation for the discussion
in Sec.~\ref{sec_results}, we consider selected limiting cases.
For the eigenstates with relative energy $E_{\rm{rel}}$
around $\hbar \bar{\omega}/2$ and
$3\hbar \bar{\omega}/2$, 
we find, respectively,
\begin{equation}
\frac{C_{\rm{E}_{\rm{rel}} \approx \hbar \bar{\omega}/2}^{\rm{adia}}}{\hbar^2 \bar{a}^{-3}} =
\frac{1}{\sqrt{2 \pi }} \left(\frac{g}{\hbar \bar{\omega} \bar{a}}\right)^2
+ \mathcal{O} \left(g^3\right)
\end{equation}
and
\begin{align}
\frac{C_{\rm{E}_{\rm{rel}} \approx 3 \hbar \bar{\omega}/2}^{\rm{adia}}}{\hbar^2 \bar{a}^{-3}} = 
2 \sqrt{\frac{2}{\pi }}
+ c_1  \frac{a_{\text{1D}}}{\bar{a}}
+ c_2  \left(\frac{a_{\text{1D}}}{\bar{a}}\right)^2
+ \mathcal{O} \left(a_{\text{1D}}^3\right),
\end{align}
where $c_1 = 0.781394\dots$ and $c_2 =-2.34671\dots$
(the analytical expressions for $c_1$ and $c_2$ are lengthy and not given here).
The one-dimensional scattering length
$a_{\rm{1D}}$ is inversely proportional to the 
one-dimensional coupling constant $g$~\cite{Olshanii98},
$a_{\rm{1D}} = -2 \hbar^2 /(m g)$.
If we assume that $g$ is small and that we
start 
in the ground state at $t=t_0$
and then change the trapping frequency adiabatically,
the ratio of the adiabatic contact at time $t$ and that at time $t_0$
is $(t_0/t)^{1/2}$.
If, on the other hand, we assume that $g$ is large ($g$ positive)
and that we start
in the ground state
at $t=t_0$
and then change the trapping frequency adiabatically,
the ratio of the adiabatic contact at time $t$ and that at time $t_0$
is $(t_0/t)^{3/2}$.
As mentioned earlier, we are primarily interested in the regime
where the Hamiltonian is changed non-adiabatically.
Thus, we expect that the time dependence of the contact is not
described accurately by the adiabatic
prescription. Instead, we expect---taking
into account that $\langle z^2(t) \rangle$
displays log-periodic oscillations
``on top'' of an overall growth in the $g=0$ and $=\infty$ limits---that
the time variation of the contact for systems with
finite $g$ will oscillate around the
adiabatic value.

It is also interesting to consider the strongly-attractive 
limit ($|g| \gg \hbar \bar{\omega} \bar{a}$ and $g < 0$),
in which
the two particles form a tightly bound dimer 
of size $a_{\text{1D}}$ ($a_{\text{1D}} \ll \bar{a}$).
In this case, the relative wave function approaches that of
two particles in free space with contact
$C_{\rm{dimer}}^{\rm{free-space}}$,
\begin{equation}
C_{\rm{dimer}}^{\rm{free-space}}= 
\frac{4 \hbar^2}{a_{\text{1D}}^3}.
\end{equation}
Since the contact
$C_{\rm{dimer}}^{\rm{free-space}}$
is independent of the harmonic oscillator length $\bar{a}$,
we expect that the time-dependence of the trapping potential for $t>t_0$
has little impact on the initial state,
provided
$a_{\text{1D}} \ll \sqrt{\hbar/(m \Omega)}$.
This is essentially saying that the trap is too weak in this limit to affect the initial
state.

\section{Two-Particle Results}
\label{sec_results}

To solve the time-dependent Schr\"odinger equation 
in the relative coordinate $z$
for the Hamiltonian with
time varying confining potential and  finite $g$,
we perform numerical calculations.
We use a propagator that exactly accounts for the two-body zero-range
interaction~\cite{Blinder88, Wodkiewicz91, Yan15}.
In brief, the relative 
coordinate is discretized using an equidistant grid
with grid spacing
$\Delta z$. 
Knowing the wave packet at time $t$,
the wave packet at time $t + \Delta t$ is obtained by integrating
the product of the propagator and the wave packet at time $t$ 
over the relative coordinate. This propagation step is repeated
for many time steps $\Delta t$.
The accuracy of the final wave packet 
depends on the values of $\Delta z$ and $\Delta t$;
implementation details can be found in
Ref.~\cite{Gharashi15-tunn}.
The errorbars (not shown) 
of the numerical
results presented in this section are smaller than the symbol
sizes or not visible
on the chosen scales.
Typical simulation parameters are 
$\Delta z = 0.003 a_{\text{ho}}(t_0)$ and 
$\Delta t = 0.04 \Omega^{-1}$.

\begin{figure}[htbp]
\centering
\includegraphics[angle=0,width=0.4\textwidth, clip=true]{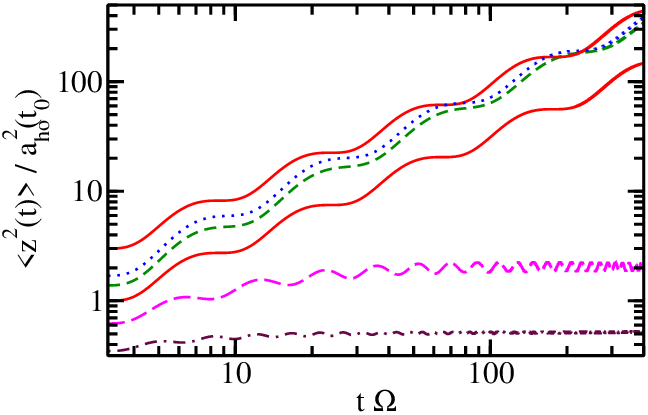}
\caption{(Color online)
Expectation value $\langle z^2(t) \rangle$ as a function of time
for two one-dimensional particles with
$\lambda = 1/10$ and various interaction strengths $g$.
The time evolution starts at $t \Omega=t_0 \Omega = \sqrt{10}$.
The lines from the top to the bottom at $t \Omega=\sqrt{10}$ 
correspond to 
$g / [E_{\rm{ho}}(t_0) a_{\rm{ho}}(t_0)] = \infty, 2, 1, 0, -1$, and $-2$.
}
\label{fig_2}
\end{figure}

Figure~\ref{fig_2}
shows the expectation value $\langle z^2(t) \rangle$ as a 
function of time for various $g$ on a log-log scale.
In all cases, the initial state at $t=t_0$ corresponds to the
lowest energy eigenstate of the time-independent $t<t_0$
Hamiltonian.
As a reference,
the two solid lines show $\langle z^2(t) \rangle$
for the systems
with scale-invariant interactions,
$|g| = \infty$ 
(upper solid curve) and $g=0$ (lower solid curve).
As discussed in Sec.~\ref{sec_scale-inv},
the two solid curves would collapse to a single curve if 
the two
$\langle z^2(t) \rangle$ were scaled 
by their corresponding initial $\langle z^2(t_0) \rangle$.
In the representation chosen in Fig.~\ref{fig_2},
the solid lines are offset from each other but
exhibit the same oscillation amplitude and period.
The dotted and short-dashed lines show the 
expectation value $\langle z^2(t) \rangle$
for repulsively-interacting systems with 
$g / [E_{\rm{ho}}(t_0) a_{\rm{ho}}(t_0)] = 2$ and 1, respectively.
For these two finite $g$ cases, 
the amplitude and oscillation period ``dephase'' over time;
this dephasing is due to the finite length scale defined by the
interactions.
Ignoring the oscillations, $\langle z^2(t) \rangle$ increases faster
for finite $g$ 
than for $g=0$.
Using a hand waving argument, this can be understood
by noticing that 
$g / [E_{\rm{ho}}(t) a_{\rm{ho}}(t)]$ increases with increasing time 
due to the time-dependence of the trapping frequency.
At large $t$, the interactions thus effectively
approach the strongly-interacting limit, explaining why the 
expectation value
$\langle z^2(t) \rangle$ for finite $g$ is closer to that
for $1/g=0$ than that for $g=0$;
a more quantitative discussion is presented below.

The long-dashed and dash-dotted lines in Fig.~\ref{fig_2}
show $\langle z^2(t) \rangle$ for
negative
interaction strengths, i.e., for
$g / [E_{\rm{ho}}(t_0) a_{\rm{ho}}(t_0)] = -1$ and $-2$, respectively.
For these $g$, $\langle z^2(t) \rangle$
first increases notably with increasing $t$ and then plateaus.
This can be intuitively understood by realizing that
the wave packet initially, oscillations aside, expands
together with the trap. At later times, however, $a_{\text{ho}}(t)$
is much larger than the size of the wave packet,
and the dynamics is approximately independent of the time
dependence of the trap, implying that
$\langle z^2(t) \rangle$ approaches 
$a_{\text{1D}}^2 /2$, i.e., the expectation value
of $z^2$
for a dimer with one-dimensional scattering length $a_{\text{1D}}$
in free space.
Indeed, this is what we observe in Fig.~\ref{fig_2}:
The long-dashed and dash-dotted lines approach $2 a_{\text{ho}}^2(t_0)$
and
$a_{\text{ho}}^2(t_0)/2$, respectively, at large $t$.
We find that 
the small oscillations exhibited by $\langle z^2(t) \rangle$ 
occur on a time scale that is, roughly, set by the two-body binding energy.

\begin{figure}[htbp]
\centering
\includegraphics[angle=0,width=0.4\textwidth, clip=true]{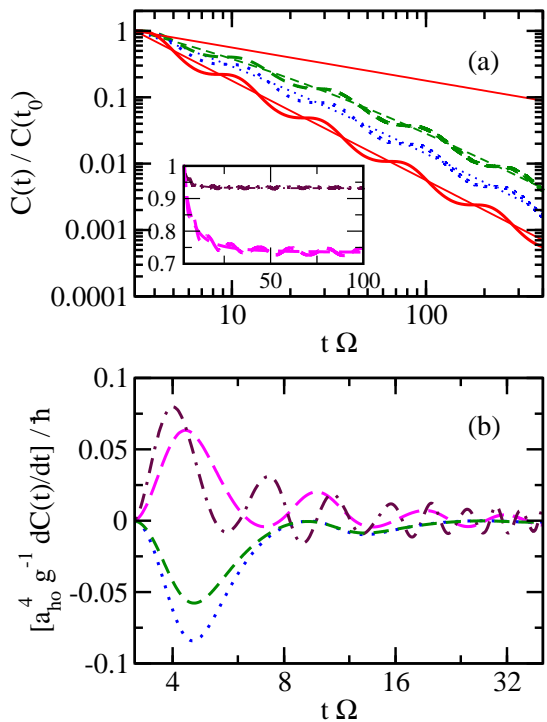}
\caption{(Color online)
Properties of the contact as a function of time 
for two one-dimensional particles with $\lambda=1/10$ and
various interaction strengths $g$.
(a) The thick lines from top to bottom show the
normalized contact $C(t) / C(t_0)$ for
$g / [E_{\rm{ho}}(t_0) a_{\rm{ho}}(t_0)] =1$
(dashed line), 
$2$
(dotted line), 
and $\infty$
(solid line),
respectively.
The thin lines show the corresponding adiabatic
contact
$C^{\text{adia}}(t)/C(t_0)$ using
the same linestyles.
The thin solid line at the top shows the limiting behavior
$\sqrt{t_0/t}$ (see text for details).
The inset shows the same quantities but for
$g / [E_{\rm{ho}}(t_0) a_{\rm{ho}}(t_0)] =-1$ (dashed lines;
upper set of curves) and 
$g / [E_{\rm{ho}}(t_0) a_{\rm{ho}}(t_0)] =-2$ 
(dash-dotted lines;
lower set of curves).
Note that the inset uses linear horizontal and vertical
scales while the main figure uses logarithmic
horizontal and vertical scales. 
(b) Dotted, short-dashed, long-dashed, and dash-dotted lines show 
the quantity $g^{-1}dC(t)/dt$
for
$g / [E_{\rm{ho}}(t_0) a_{\rm{ho}}(t_0)] =2$, $1$, $-1$, and $-2$,
respectively.
}
\label{fig_3}
\end{figure}

As discussed in Sec.~\ref{sec_broken_scale-inv},
the length scale introduced by the two-body interaction
manifests itself
in the differential equation
for 
$\langle z^2(t) \rangle$ via the time derivative $d C(t)/dt$ of the
contact [see Eq.~(\ref{Eq_diff_Zhai})].
To analyze the role of this ``new'' term,
thick lines in Fig.~\ref{fig_3}(a)
show the contact $C(t)$, 
normalized by its initial value $C(t_0)$, for $\lambda = 1/10$ and 
various $g$.
For finite $g$, $C(t)$ displays oscillations on top of an
overall decay.
The overall decay is well described by the
adiabatic contact $C^{\text{adia}}(t)$,
which is shown by the thin lines for each $g$ considered.
As already discussed in Sec.~\ref{sec_broken_scale-inv},
 the thin lines for the strongly-repulsive systems fall
off approximately as $(t_0/t)^{3/2}$ for all $t$ 
(the approximation becomes better for larger $t$)
while the thin lines for the weakly-repulsive systems
transition from an initial $(t_0/t)^{1/2}$ fall-off 
to a $(t_0/t)^{3/2}$ fall-off for large $t$.
The non-trivial time-dependence of $C(t)$ 
is responsible for the ``dephasing''
discussed in the context of Fig.~\ref{fig_2}.

To see this more clearly,
Fig.~\ref{fig_3}(b) shows
the quantity
$g^{-1}dC(t)/dt$, i.e., the right hand side of Eq.~(\ref{Eq_diff_1D}),
for various finite $g$.
While this quantity vanishes for $g = 0$ and $g = \infty$, it
exhibits oscillations on top
of an overall decrease with increasing $t$ for finite $g$.
Our analysis shows that the absolute value of each of the three
terms on the left hand side of equation Eq.~(\ref{Eq_diff_1D})
is, for the first few oscillations, 
of the same order of magnitude as the absolute 
value of the right hand side of that equation. 
As time increases, the magnitude of the 
right hand side of Eq.~(\ref{Eq_diff_1D}),
however,
decreases faster than that of the other three terms. 
For positive $g$, we find that the relative importance of
the right hand side decreases faster for larger $g$.
This can, again, be intuitively understood by realizing that
$|g|/[E_{\text{ho}}(t) a_{\text{ho}}(t)]$ increases with increasing time.

To analyze the time dependence 
of
$\langle z^2(t) \rangle$ 
further,
we think of the dynamics displayed in Fig.~\ref{fig_2}
for finite $g$ as 
``close to periodic'' and devise empirical measures to
quantify the deviations from the
``truly periodic'' dynamics encountered for
$g=0$ and $1/g=0$.
To this end, we calculate $d\langle z^2 (t) \rangle/dt$.
While 
$d\langle z^2 (t) \rangle/dt$
is zero at the beginning and end of each cycle for $g=0$ or $1/g=0$ 
(see Fig.~\ref{fig_phasespace}),
it does not go to zero for finite $g$.
For positive $g$,
we thus define cycles by looking at
consecutive local minima 
of $d\langle z^2 (t) \rangle/dt$
and 
by
denoting the time at which the $j$th cycle starts
by $t_{\rm{in}}^{(j)}$ 
and the time 
at which the $j$th
cycle ends by $t_{\rm{fi}}^{(j)}$
(the cycle enumeration starts with $j=1$).
We define
\begin{eqnarray}
c^{(j)}=\frac{d \langle z^2(t) \rangle}{dt} \Big| _{t=t_{\rm{fi}}^{(j)}}
\label{Eq_slope}
\end{eqnarray}
as well as the scaling factors $\tau^{(j)}$ and $\zeta^{(j)}$,
\begin{equation} 
\tau^{(j)} =
{t_{\rm{fi}}^{(j)}}/
{t_{\rm{in}}^{(j)}},
\label{Eq_tau}
\end{equation}
and
\begin{equation} 
\zeta^{(j)} = 
\frac
{\langle z^2 (t_{\rm{fi}}^{(j)}) \rangle}
{\langle z^2 (t_{\rm{in}}^{(j)}) \rangle}.
\label{Eq_zeta}
\end{equation}
For $g=0$ and $|g|=\infty$,
these scaling factors are independent of $j$
and given by
$\tau= \zeta = \exp(2 \pi / s_0)$.

\begin{figure}[htbp]
\centering
\includegraphics[angle=0,width=0.4\textwidth, clip=true]{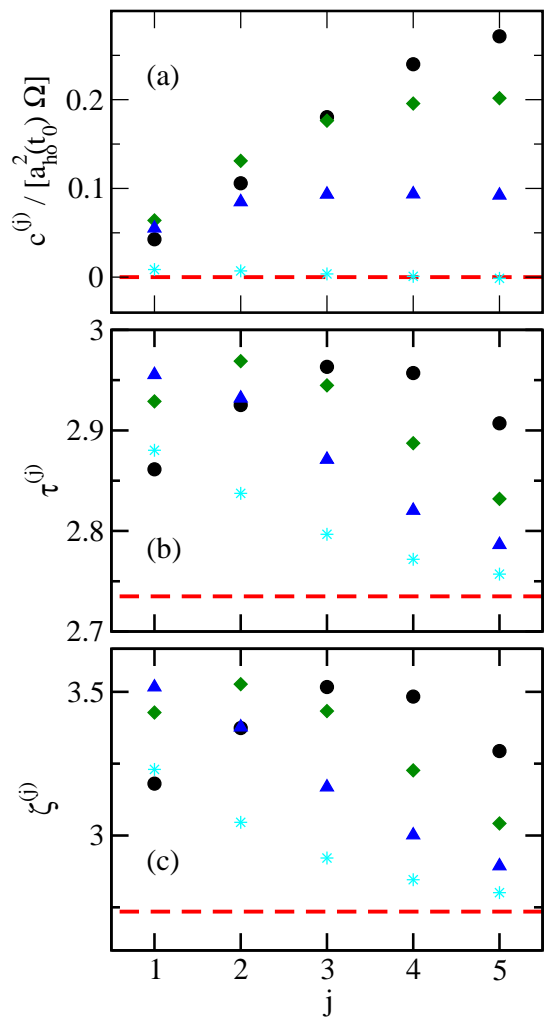}
\caption{(Color online)
Characterization of the
expansion dynamics as a function
of the cycle number $j$ for
two-body
systems with $\lambda = 1/10$ and positive $g$.
The symbols in panels~(a), (b), and (c) show
$c^{(j)}$ [Eq.~(\ref{Eq_slope})],  
$\tau^{(j)}$ [Eq.~(\ref{Eq_tau})], and 
$\zeta^{(j)}$ [Eq.~(\ref{Eq_zeta})], respectively.
The circles, diamonds, triangles, and stars correspond to 
$g / [E_{\rm{ho}}(t_0) a_{\rm{ho}}(t_0)] = 1/2, 1,2$, and $5$,
respectively.
The horizontal dashed lines show the corresponding
values for the case where the two-body interaction
does not define a meaningful length scale ($g=0$ or $|g|=\infty$);
in this case, we have
$\tau = \zeta=\exp( 2 \pi / s_0) \approx 2.734955$.
}
\label{fig_4}
\end{figure} 

Circles, diamonds, triangles, and stars in Fig.~\ref{fig_4}
show our numerical results
for Eqs.~(\ref{Eq_slope})-(\ref{Eq_zeta})
for 
$g / [E_{\rm{ho}}(t_0) a_{\rm{ho}}(t_0)] = 1/2, 1, 2$, and $5$, respectively.
For comparison,
the dashed lines show the corresponding values for the
scale-invariant systems.
For the two largest $g$ considered (stars and triangles),
$\tau^{(j)}$ and $\zeta^{(j)}$ decrease monotonically with increasing
$j$. Figures~\ref{fig_4}(b) and \ref{fig_4}(c) suggest that
the large $j$ limit is given by the horizontal dashed lines, 
i.e., that the 
scale factors for finite $g$ 
at large $t$ 
approach the scale factor of the scale-invariant systems.
For smaller $g$ (diamonds and circles), the scale factors first 
increase, then reach a maximum, and finally decrease.
Although our numerics does not allow us to go beyond $j=5$,
Figs.~\ref{fig_4}(b) and \ref{fig_4}(c) suggest that the 
scale factors for systems with these smaller $g$ values
also approach the scale-invariant value in the large $j$ limit.
The quantity $c^{(j)}$ [see Fig.~\ref{fig_4}(a)], in contrast,
does not approach the value of the scale invariant systems but instead approaches a finite
constant for large $j$. The asymptotic, large $j$ value is reached
faster for large $g$ than for small $g$.

We now present analytical considerations for 
systems with finite positive $g$ that explain the 
trends displayed in Fig.~\ref{fig_4}.
As mentioned already several times, 
$g/[E_{\text{ho}}(t) a_{\text{ho}}(t)]$
increases with increasing time. 
Thus,
we neglect the right hand side of Eq.~(\ref{Eq_diff_1D})
in the large $t$ limit and 
analyze the analytical solution,
which can be found in Ref.~\cite{Lewis68},
 for the differential equation
assuming finite values for $\langle z^2(t') \rangle$ and
$[d \langle z^2(t) \rangle/dt]_{t=t'}$, where $t' \gg t_0$.
The reason for using a finite value for
$[d \langle z^2(t) \rangle/dt]_{t=t'}$ is that the wave packet at 
$t=t'$ is not in an eigenstate [indeed,
Fig.~\ref{fig_4}(a) shows that $c^{(j)}$ 
is finite].
We find that the
analytical solution for the initial
conditions applicable to an initial non-stationary state
exhibits the same
log-periodic oscillation period and amplitude
as the analytical solution for the initial
conditions applicable
to an initial stationary state.
This explains why the scaling 
factors $\tau^{(j)}$ and $\zeta^{(j)}$ [see Figs.~\ref{fig_4}(b)
and \ref{fig_4}(c)] approach
the dashed horizontal lines in the large $j$ limit.
Moreover, since the analytical description is expected to become more
accurate for larger $j$,
$c^{(j)}$ is expected to approach a constant
in the large $j$ limit.

\begin{figure}[htbp]
\centering
\includegraphics[angle=0,width=0.4\textwidth, clip=true]{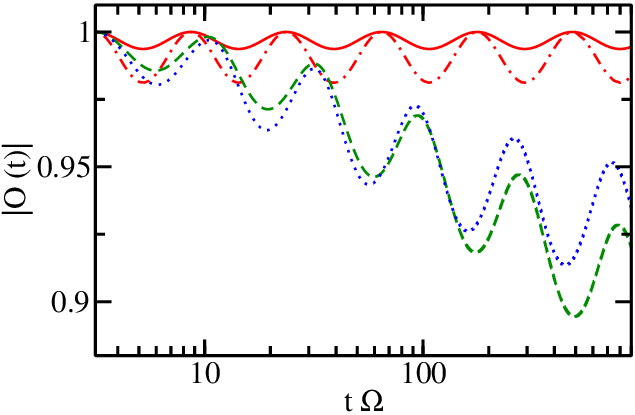}
\caption{(Color online)
Norm $|O(t)|$ of the overlap [see Eq.~(\ref{eq_overlap})]
as a function of time
for two one-dimensional particles with $\lambda=1/10$
and various interaction strengths $g$.
Solid, dash-dotted, dashed, and dotted lines correspond to
$g / [E_{\rm{ho}}(t_0) a_{\rm{ho}}(t_0)] = 0$, 
$\infty$, $1$, and $2$, respectively.
}
\label{fig_5}
\end{figure}

While the discussion above shows that certain properties
of the large $t$ dynamics can be predicted analytically, 
the full wave packet dynamics for finite $g$ may
be quite intricate, 
i.e.,
the quantities $\langle z^2(t) \rangle$
and
$d \langle z^2(t) \rangle/dt$
alone may not describe the full story.
To corroborate this notion,
we consider
the time-dependent overlap $O(t)$ 
between the wave packet $\psi(z,t)$ and the adiabatic eigenstate 
$\psi^{\rm{adia}}(z, a_{\rm{ho}}(t))$ for the same $g$
and confinement with harmonic oscillator length $a_{\text{ho}}(t)$,
\begin{equation}
\label{eq_overlap}
O(t) = 
\langle \psi(z,t) | 
\psi^{\rm{adia}}\left(z, a_{\rm{ho}}(t)\right) \rangle.
\end{equation}
Figure~\ref{fig_5}
shows $|O(t)|$ 
(the absolute value is taken to eliminate
arbitrary phase factors)
as a function of time for various coupling strengths $g$. 
For $g = 0$ (solid line)
and $\infty$ (dash-dotted line),
$|O(t)|$ is equal to 1 at the end of each cycle,
reflecting the fact that the wave packet returns,
provided the appropriate scaling of the $z$ coordinate is applied, 
to its orginal shape.
Interestingly, $|O(t)|$ takes on its minimal values
at the ``half-cycle'' times, i.e., at the times 
where $\langle z^2(t) \rangle$ 
coincides
with the corresponding adiabatic value.
For finite $g$, the norm $|O(t)|$ of the overlap
displays, as for the scale-invariant interactions,
oscillatory behavior. However, the oscillations are
on top of an overall decrease of the overlap.
This implies that the finite length scale continually
introduces a dephasing, i.e., the time-dependent 
wave packet is increasingly less similar to the adiabatic
eigenstate. Said differently, excited adiabatic states
get mixed in more with increasing time $t$.
Taking a slightly different view point,
this means that a full description of the wave packet at the
end of the empirically defined cycles requires
knowledge not only of
$\langle z^2(t) \rangle$ but  
of $\langle z^n(t) \rangle$ 
for all $n$.

\section{Conclusion}
\label{sec_conclusion}

This paper investigated the expansion dynamics of
a harmonically trapped 
cold atom system with zero-range interactions.
For $t<t_0$, the
angular
trap frequency $\omega(t)$ is equal to the
constant $\Omega$
and the system is in an eigenstate.
For $t>t_0$, a scale-invariant trap
potential is realized by varying the angular frequency 
smoothly according to $\omega(t) =\Omega t_0/t$.
The scale invariance can be most
readily seen from Eq.~(\ref{Eq_diff_Zhai}), where $R$
and $t$ occur with the same powers in
all three terms.
More formally, we can look at how the Hamiltonian
changes if
the position coordinates
are multiplied by $\alpha$. In this case,
the kinetic energy term picks up an extra $\alpha^{-2}$ 
factor 
while the time-independent trapping potential
picks up an extra factor of $\alpha^2$; thus, the
system Hamiltonian possesses a scale
(namely the harmonic oscillator length). Scaling the time by
$\alpha^2$ (this follows from the fact that time
is, dimensionally, 
$\hbar$ divided by energy and that energy scales as 
one over length to the power of two),
the time-dependent trapping potential, which
contains terms like 
$\omega^2(t) \mathbf{r}_j^2$,
picks up a factor of $\alpha^{-2}$, just as the kinetic energy term;
thus, the system Hamiltonian does not possess a scale,
i.e., it is scale-invariant.

If the interaction strength $g$ vanishes or is infinitely
large, the entire systems Hamiltonian
is scale-invariant. In this case, the expansion dynamics has been
investigated experimentally and theoretically in Ref.~\cite{Deng16}.
It was found that the cloud size follows so-called Efimovian 
expansion dynamics with logarithmically spaced
oscillation periods and amplitudes.

The present paper investigated how the expansion dynamics changes
if the interaction strength $g$ is finite.
To address this question, the simplest non-trivial system
consisting of two one-dimensional
particles was investigated. It was found that the 
expansion dynamics of 
systems with finite and positive $g$ still exhibits 
oscillatory behavior. However, the oscillation period and amplitude
are no longer logarithmically spaced. Instead, a dephasing that 
is governed by the time derivative of the contact is observed.
At large times (depending on the value
of $g$, this may imply large numbers of ``close-to-periodic''
oscillations),
we found that the cloud size is again, at least approximately, governed 
by a unique oscillation period and amplitude.
We note that
the role of the contact in non-equilibrium situations was previously
investigated in two-atom quantum quenches~\cite{Corson16}, where 
a rapid change of the interaction strength (a quench) induced a ballistic 
component into the contact.
In the scenario considered in the present paper,
the oscillations of the contact are the result
of the continuously changing trapping potential. The work done
on the system for $t>t_0$ triggers an interplay between
the harmonic oscillator parts of the Hamiltonian and 
the two-body interaction terms. When the cloud is extremely dilute
(large times), the effective interaction strength
of the one-dimensional system becomes large 
[$g/[E_{\text{ho}}(t)a_{\text{ho}}(t)]$
increases as $g \sqrt{t}$].
Thus, the system is, again, effectively scale-invariant at large times,
implying---using the general solutions
for the scale-invariant system with modified
initial conditions---close-to-log-periodic expansion dynamics.
While our analysis in Sec.~\ref{sec_results}
considered two one-dimensional particles with finite interaction
strength, the results generalize to systems consisting 
of more particles
and with other dimensionalities.

The design of time-dependent trapping potentials has played an
important role over the past 60 years or so.
Here, the role of finite
two-body interaction
strengths
on the Efimovian-like expansion dynamics 
was investigated. In other contexts 
(see, e.g., Ref.~\cite{Chen10}), time-dependent
trapping potentials have been used to design
frictionless non-adiabatic atom cooling trajectories. It
would be interesting to include
atom-atom interactions
in those contexts via the contact and to quantify
the energy distribution among the various modes,
building on the ideas put forward in Ref.~\cite{Ebert15}.

\section{Acknowledgments}
\label{sec_acknow}
Support by the National
Science Foundation (NSF) through Grant No.
PHY-1415112
is gratefully acknowledged.
This work used the Extreme Science and Engineering
Discovery Environment (XSEDE), which is supported by
NSF Grant No. OCI-1053575, and the
WSU HPC.



\begin{thebibliography}{51}
\bibitem [{footnote7-1()}]{footnote7-1}%
  \BibitemOpen
  \bibinfo {note} {An example is the two-component Fermi gas with vanishing
  intraspecies interaction strength and infinitely large interspecies
  interaction strength.}\BibitemShut {Stop}%
\end{thebibliography}
%

\end{document}